\documentclass[final]{article}

\usepackage{neurips_2022}

\usepackage[utf8]{inputenc} 
\usepackage[T1]{fontenc}    
\usepackage{hyperref}       
\usepackage{url}            
\usepackage{booktabs}       
\usepackage{amsfonts}       
\usepackage{nicefrac}       
\usepackage{microtype}      
\usepackage[table,xcdraw]{xcolor}

\usepackage{graphicx}
\usepackage{caption}
\usepackage{subcaption}
\usepackage{grffile}

\newcommand*\samethanks[1][\value{footnote}]{\footnotemark[#1]}

\title{EN: Good Practices for Institutional Organization of Research Institutes: Excellence in Research and Positive Impact on Society\\
BOS: Dobre prakse institucionalne organizacije istraživačkih instituta: izvrsnost u istraživanju i pozitivan uticaj na društvo}

\author{%
  Zlatan Ajanović\thanks{members of the Association for Advancement Science and Technology (ANNT), Bosnia and Herzegovina.} \\
  RWTH Aachen University\\ 
  Aachen, Germany\\
  \texttt{zlatan.ajanovic@rwth-aachen.de} 
  \And
  Hamza Merzić\samethanks\\
  Google DeepMind\\
  London, UK
  \And
  Suad Krilasević\samethanks\\
  Asocijacija za napredak nauke i tehnologije\\
  Sarajevo, Bosnia and Hercegovina
  \And
  Eldar Kurtić\samethanks \\
  IST Austria \\
  Vienna, Austria
  \And
  Bakir Kudić\samethanks \\
  Francis Crick Institute, \\
  London, UK
  \And
  Rialda Spahić\samethanks \\
  Equinor\\
  Stavanger, Norway
  \And
  Emina Aličković\samethanks \\
  Linköping University\\
  Linköping, Sweden
  \And
  Aida Branković\samethanks \\
  CSIRO, Health Intelligence \& \\
  The University of Queensland \\
  Queensland, Australia
  \And
  Kenan Šehić\samethanks \\
  University of Sarajevo\\
  Sarajevo, Bosnia and Herzegovina 
  \And
  Mirsad Ćosović\samethanks\\
  University of Sarajevo\\
  Sarajevo, Bosnia and Herzegovina
  \And
  Admir Greljo\samethanks\\
  University of Basel\\
  Basel, Switzerland
  \And
  Sead Delalić\samethanks \\
  University of Sarajevo\\
  Sarajevo, Bosnia and Herzegovina
  \And
  Adnan Mehonić\samethanks \\
  University College London \\
  London, UK 
}

\begin{document}

\maketitle

\begin{abstract}
\textbf{EN:}
In this paper, we analyze examples of research institutes that stand out in scientific excellence and social impact. We define key practices for evaluating research results, economic conditions, and the selection of specific research topics. Special focus is placed on small countries and the field of artificial intelligence. The aim is to identify components that enable institutes to achieve a high level of innovation, self-sustainability, and social benefits.

\textbf{BOS:}
U ovom radu analiziramo primjere istraživačkih instituta, a koji se ističu u naučnoj izvrsnosti i društvenom utjecaju. Definiramo ključne prakse za evaluaciju istraživačkih rezultata, ekonomske uslove i odabir specifičnih istraživačkih tema. Poseban fokus stavljamo na male zemlje i oblast vještačke inteligencije. Cilj je identificirati strukture koje omogućuju institutima da postignu visoku razinu inovacija, samoodrživosti i društvene koristi.

\end{abstract}

\section{Uvod}
Protekle dvije godine obilježile su značajan napredak u oblasti umjetne inteligencije (AI) na globalnom nivou, ali i unutar Bosne i Hercegovine. Ovaj rad predstavlja nastavak prethodnih istraživanja i procjena razvoja AI tehnologija i njihovog utjecaja na društvo i ekonomiju, s posebnim osvrtom na lokalne inicijative, projekte i promjene u akademskim i poslovnim krugovima. U prethodnom radu (Ajanović et al. 2022), predstavljena je AI vizija za BiH, koja je obuhvatila ključne korake potrebne za implementaciju AI u javni i privatni sektor, kao i neophodnost izgradnje adekvatnih strategija, infrastrukture i kapaciteta. U međuvremenu, globalna AI scena nastavila je da se ubrzano razvija, donoseći nove tehnologije i etičke izazove, dok je Bosna i Hercegovina postepeno ulagala u razvoj obrazovanja, infrastrukture i istraživačkih projekata vezanih za AI.

Ovaj rad je pisan s ciljem da bude pristupačan široj publici, uključujući i donosioce odluka, te nudi relevantne reference za dalja istraživanja. Fokus je na praktičnim primjerima i dobrim praksama izvučenim iz iskustava autora i stručnjaka u ovoj oblasti. Rad se bavi temama poput napora u obrazovanju, institucionalnoj podršci i digitalizaciji javnih usluga, kao i razvoju etičkih okvira i regulacija. Također, pruža osvrt na organizaciju naučno-istraživačkog rada, s naglaskom na interdisciplinarnost i inovacije.

Struktura rada je organizovana na način da omogućava jasno razumijevanje trenutnih trendova i izazova u AI oblasti. Prvo se daje pregled globalnih i lokalnih dešavanja u protekle dvije godine u AI sferi. Zatim se analizira napredak u edukaciji, sa posebnim fokusom na nove studijske programe i inicijative. U nastavku se obrađuju tehnološka dostignuća i njihove primjene, kao i etički i regulativni aspekti, uz osvrt na standarde i javne politike. Poseban dio posvećen je organizaciji naučno-istraživačkog rada, uključujući pregled ključnih preduslova za uspješan razvoj istraživanja u AI oblasti u Bosni i Hercegovini. Rad završava preporukama za dalji razvoj AI infrastrukture i strategija na nacionalnom nivou.

\section{Dvije godine od AI vizije}

U protekle dvije do tri godine, globalni tehnološki napredak u umjetnoj inteligenciji ubrzano se razvijao, uz naglasak na etiku, regulaciju i standardizaciju. Bosna i Hercegovina počinje pratiti ove trendove, posebno kroz razvoj obrazovnih programa i digitalizaciju. Ovo poglavlje daje pregled ključnih događaja u tom periodu, uključujući tehnološki napredak, jačanje AI institucija i održavanje prvih domaćih AI konferencija, kao nastavak istraživanja predstavljenog u prethodnom radu.

\subsection{Obrazovanje}

Umjetna inteligencija godinama postaje sve značajnija u bh. društvu, a obrazovne institucije prepoznaju potrebu za formalnim obrazovanjem u ovoj oblasti. Tokom ranijih godina, razvoj AI zajednice u BiH značajno se oslanjao na pojedince koji su svoje obrazovanje stekli na Univerzitetu u Sarajevu (UNSA), ali su značajan dio vremena proveli u samostalnoj edukaciji o ovoj oblasti. Studij Kompjuterskih nauka na Prirodno-matematičkom fakultetu (PMF) i srodni studij na Odsjeku za računarstvo i informatiku pri Elektrotehničkom fakultetu (ETF) pružaju kvalitetno opšte znanje u računarstvu, što predstavlja bitnu osnovu za razvoj AI inženjera. Stalnim promjenama i modernizacijom studijskih programa i kurseva, studenti na višim godinama imaju priliku izučavati specijalizovane predmete, poput Uvoda u vještačku inteligenciju, Mašinskog učenja, Neuralnih mreža, Kompjuterske vizije i drugih AI disciplina. Naravno, stručnost pojedinaca i dalje često proizlazi iz lične posvećenosti vještačkoj inteligenciji, te dodatnom trudu uloženom izvan nastave, kao i kroz angažman na AI projektima. Dio AI stručnjaka u Bosni i Hercegovini je završio studij i na Odsjeku za automatiku i elektroniku pri ETF-u, koji je prepoznatljiv po kvalitetnim studentima, iako nije primarno fokusiran na AI ili računarstvo.

Za dodatni rast AI zajednice u BiH, potrebni su i specijalizirani studiji koji će omogućiti studentima da se od prvog dana usmjere na AI. Elektrotehnički fakultet UNSA je među prvima u regionu koji pokreće studijski program usmjeren na AI i Data Science (Univerzitet u Sarajevu 2024), inspirisan sličnim programima u zemljama Europske unije (EU). Ova specijalizirana univerzitetska diploma omogućit će studentima da budu konkurentni na tržištu rada u različitim domenama primjene AI, jer primjena AI nadilazi tehničke nauke i nalazi primjenu u oblastima poput medicine, farmacije, proizvodnog inženjeringa i mnogih drugih. Tokom trogodišnjeg studija, koncipiranog na engleskom jeziku, studenti će imati priliku da detaljno izučavaju specijalizirane predmete iz oblasti AI, od matematike koja se koristi u AI do praktične primjene AI u raznim oblastima. Iako je Internacionalni Univerzitet u Sarajevu također pokrenuo inicijativu za stvaranje studijskog programa usmjerenog na AI i Data Engineering (obrada podataka), analizom silabusa (Internacionalni Univerzitet u Sarajevu 2023) možemo primijetiti da su predmeti ipak sličnog koncepta kao opšti studijski program na ETF-u na Odsjeku za računarstvo i informatiku ili studiju Kompjuterskih nauka na PMF-u. 
S druge strane, važno je naglasiti i potencijalnu opasnost specijaliziranih studijskih programa, jer u slučaju nedostatka radnih mjesta i AI projekata u IT kompanijama, studenti nemaju dovoljno opšteg znanja za obavljanje drugih poslova, poput razvoja Web sistema, što je i dalje osnova poslovanja većina IT kompanija u BiH. Dodatno, svjedoci smo da je bh. IT zajednica intenzivirala organizaciju različitih AI radionica, gdje pojedinci iz raznih polja mogu steći prvo iskustvo u radu s AI-om. Google Developer Group Sarajevo kroz svoje radionice upoznaje učesnike s kapacitetima Google Clouda za primjenu AI u različitim aplikacijama. Većina ovih radionica je besplatna i predstavlja odličnu priliku za pojedince koji žele saznati više o vještačkoj inteligenciji i njenim potencijalima. Studentske organizacije u saradnji s industrijom intenzivno organizuju "hackathone" gdje studenti provode dane rješavajući izazove koji zahtijevaju kreativna rješenja koristeći AI u svakodnevnom životu. Ovi hackathoni pružaju studentima priliku da primijene svoje znanje u praktičnim situacijama, radeći na projektima koji mogu imati stvaran uticaj na društvo. Na primjer, tokom ovogodišnjeg hackathona EESTech Challenge, studenti su razvijali inovativna rješenja za upotrebu AI sa pametnim mjeračima u domaćinstvu. Također, u organizaciji Asocijacije za napredak nauke i tehnologije, Bosanskohercegovačko-američke akademije umjetnosti i nauka i RunIT organizovan je AI Hackathon sa ciljem korištenja velikih jezičkih modela za izradu dodatnih funkcionalnosti na Akademskom Imeniku\footnote{https://akademskiimenik.ba/}. Ovi događaji ne samo da podstiču inovativnost i timski rad, već i omogućavaju studentima da se povežu s mentorima iz industrije i potencijalnim poslodavcima, čime dodatno jačaju svoje profesionalne mreže.

Udruženje matematičara Kantona Sarajevo je u prethodnim godinama organizovalo više edukacija i kurseva za učenike srednjih, pa čak i osnovnih škola koji su vezani za primjenu i razumijevanje AI algoritama, poput kurseva u Školi programiranja. Studentske organizacije STELEKS, Udruženje Plus i drugi, kroz svoje događaje, druženja i AI radionice organizuju niz predavanja iz ove oblasti. Kompanije poput Revicon-a, koje organizuju edukacije za stručnjake iz raznih oblasti (ekonomija, pravo i drugo), počele su uključivati AI edukacije u svoju ponudu.
Važno je naglasiti da je Internet promijenio pristup obrazovanju, omogućavajući pojedincima da lahko pristupe online kursevima i programima. Platforme kao što su Udacity, Coursera, edX i mnoge druge nude širok spektar kurseva koji pokrivaju različite aspekte vještačke inteligencije. Ovi kursevi nisu samo tehnički, već su prilagođeni i za donosioce odluka, menadžere i profesionalce iz različitih oblasti. Dolaskom AI alata kao što je ChatGPT u obrazovanje, došlo je do značajnih poboljšanja u iskustvu učenja i podučavanja. ChatGPT može djelovati kao pametni besplatni tutor, pružajući brzu podršku studentima, objašnjavajući složene pojmove, odgovarajući na pitanja i nudeći dodatne resurse za učenje. Ovo pomaže učenicima da bolje razumiju gradivo i unaprijede svoje znanje. S druge strane, ovo će predstavljati izazov za nastavno osoblje, koje će morati unaprijediti zadatke kako bi osiguralo da studenti razvijaju kritičko razmišljanje i kreativne vještine, umjesto da se slijepo oslanjaju na AI. Nastavnici će morati osmisliti inovativne metode ocjenjivanja koje potiču originalnost, analitičko mišljenje i sposobnost rješavanja problema, čime će se osigurati da AI bude alat za podršku, a ne zamjena za aktivno učenje.

\subsection{AI Institucije}
Istraživanja u oblasti vještačke inteligencije u Bosni i Hercegovini još uvijek su primarno vezana za rad na univerzitetima, uz povremene inicijative domaćih IT kompanija. Najveći broj naučnih rezultata objavljuje se kroz rad profesora, asistenata i doktoranata, te kroz organizaciju istraživačkih grupa. Ovi naučni rezultati često su podržani kroz projekte, ali u velikoj mjeri i rezultat su individualnog angažmana.

Nekoliko IT kompanija u BiH je u posljednjih nekoliko godina finansiralo doktorske studije svojih zaposlenika u oblasti vještačke inteligencije. Također, podržali su niz edukacija iz ove oblasti za svoje uposlenike te pisanje i objavu naučnih i stručnih istraživanja. Među prvim kompanijama koje su prepoznale značaj vještačke inteligencije je Info Studio iz Sarajeva, koja je u prethodnim godinama finansirala objavu više od 30 naučnih radova iz oblasti AI i primjenu AI rješenja u svojim proizvodima. Kompanija Infobip omogućila je svojim AI inženjerima patentiranje rješenja i podržala objavu nekoliko naučnih radova. Ove kompanije ostvarile su blisku saradnju s fakultetima na implementaciji zajedničkih projekata, doktorskih studija, kurseva, ali i modifikaciji studijskih programa s ciljem prilagodbe industriji.

Primjetan je trend u regionu i šire pokretanja novih AI instituta, kako privatnih tako i javnih. Primjer privatnih instituta koji su osnovani u BiH u posljednjih nekoliko godina uključuju Verlab Institut i Blum Institut. Rezultati djelovanja i naučni doprinos još uvijek ne mogu biti procijenjeni. U regionu i Evropi je značajan broj institucija koje su za kratak period postigle zapažene rezultate. Neke od njih su predstavljene u nastavku.

\subsection{AI konferencije i časopisi}

Umjetna inteligencija je već dobro uspostavljena istraživačka oblast, sa brojnim konferencijama koje se održavaju više od 30 godina. Konferencije predstavljaju jedan od glavnih načina distribucije istraživanja u oblasti vještačke inteligencije i istraživači koriste ove događaje da predstave svoja najnovija otkrića, razmijene ideje i inovacije, te izgrade saradnje sa stručnjacima iz cijelog svijeta.

Kategorizacija i evaluacija AI konferencija i časopisa ključni su za procjenu kvaliteta i uticaja istraživačkih radova. Postoji nekoliko standardizovanih sistema koji se koriste za ocjenu i rangiranje konferencija i časopisa, što omogućava istraživačima i institucijama da prepoznaju najrelevantnije platforme za predstavljanje naučnih rezultata. Glavni sistemi uključuju:
\begin{itemize}
    \item SCImago Journal Rank (SJR), koji rangira časopise i konferencije na osnovu broja citata i ukupnog uticaja objavljenih radova.
    \item Web of Science, koji koristi niz pokazatelja za procjenu kvaliteta konferencija i časopisa, uključujući h-indeks i faktor uticaja.
    \item Core Ranking, koji klasifikuje konferencije od A* (najviši rang) do C (niži rang), pri čemu su najprestižnije AI konferencije poput NeurIPS, ICML, AAAI, ICAPS, CVPR i ICCV obično rangirane kao A*.
\end{itemize}

Pored standardizovanih sistema, Google Scholar je najčešće korišten alat za procjenu naučnog uticaja pojedinaca, grupa i institucija kroz metrike poput h-indeksa i citiranosti radova. Iako nije standardizovan kao SJR ili Web of Science, Google Scholar je javan i nudi praktične informacije o popularnosti i relevantnosti istraživačkih radova.

Najprestižnije svjetske AI konferencije prema temama uključuju:

\textbf{Generalno (široka primjena vještačke inteligencije):}
\begin{itemize}
    \item \textbf{NeurIPS} (Conference on Neural Information Processing Systems): Najuticajnija konferencija u oblasti neuronskih mreža, mašinskog učenja i statistike.
    \item \textbf{ICML} (International Conference on Machine Learning): Vodeća konferencija za teoriju i primjenu mašinskog učenja.
    \item \textbf{ICLR} (International Conference on Learning Representations): Fokusirana na duboko učenje i reprezentacije u učenju.
    \item \textbf{AAAI} (Association for the Advancement of Artificial Intelligence): Pokriva širok spektar AI tema, uključujući etiku i transparentnost u AI sistemima.
    \item \textbf{IJCAI} (International Joint Conference on Artificial Intelligence): Jedna od najstarijih konferencija, koja pokriva široke aspekte vještačke inteligencije.
\end{itemize}

\textbf{Kompjuterska vizija:}
\begin{itemize}
    \item \textbf{ICCV} (International Conference on Computer Vision): Najprestižnija konferencija za kompjutersku viziju, uključujući 3D rekonstrukciju i prepoznavanje objekata.
    \item \textbf{CVPR} (Conference on Computer Vision and Pattern Recognition): Fokusirana na prepoznavanje obrazaca i generativne modele u kompjuterskoj viziji.
    \item \textbf{ECCV} (European Conference on Computer Vision): Glavna europska konferencija za kompjutersku viziju.
\end{itemize}

\textbf{Obrada prirodnog jezika:}
\begin{itemize}
    \item \textbf{EMNLP} (Conference on Empirical Methods in Natural Language Processing): Konferencija posvećena empirijskim metodama u obradi prirodnog jezika.
    \item \textbf{ACL} (Annual Meeting of the Association for Computational Linguistics): Vodeća konferencija za obradu prirodnog jezika i jezičke modele.
\end{itemize}

\textbf{Simboličko zaključivanje i planiranje:}
\begin{itemize}
    \item \textbf{KR} (International Conference on Principles of Knowledge Representation and Reasoning): Konferencija koja se bavi simboličkim zaključivanjem i logičkim sistemima.
    \item \textbf{ICAPS} (International Conference on Automated Planning and Scheduling): Fokusirana na automatizovano planiranje i raspoređivanje.
\end{itemize}

\textbf{Robotika i automatizacija:}
\begin{itemize}
    \item \textbf{ICRA} (IEEE International Conference on Robotics and Automation): Vodeća globalna konferencija za robotiku i automatizaciju.
    \item \textbf{IROS} (IEEE/RSJ International Conference on Intelligent Robots and Systems): Usmjerena na inovacije u inteligentnim robotskim sistemima.
    \item \textbf{RSS} (Robotics: Science and Systems): Naučno orijentisana konferencija koja se fokusira na teorijske temelje robotike.
    \item \textbf{CoRL} (Conference on Robot Learning): Posvećena istraživanju mašinskog učenja u kontekstu robota i autonomnih sistema.
\end{itemize}

U Bosni i Hercegovini, standardi rangiranja i kategorizacije konferencija i časopisa još uvijek nisu u potpunosti primijenjeni, što istraživačima otežava pristup međunarodnim fondovima i prepoznavanje njihovog rada. Dok mnoge susjedne zemlje primjenjuju ove standarde, istraživači u Bosni i Hercegovini se suočavaju sa izazovima u ostvarivanju međunarodnog priznanja.
Iako se u Bosni i Hercegovini ne održavaju prestižne konferencije na svjetskom nivou, postoje određeni napori u organizaciji događaja posvećenih AI temama. Ovi događaji, iako manje uticajni, pružaju priliku lokalnim istraživačima da se uključe u globalne trendove i razmjene znanja. 
Primjeri uključuju:
\begin{itemize}
    \item \textbf{ICAT} (International Conference on Information, Communication and Automation Technologies), koja pokriva teme vezane za AI i okuplja govornike iz zemlje i svijeta.
    \item \textbf{„Umjetna inteligencija u Bosni i Hercegovini - istraživanje, primjena i perspektive razvoja“}, koju organizuju Federalno ministarstvo obrazovanja i nauke i INTERA Tehnološki park Mostar, sa ciljem povezivanja istraživača i razmjene znanja.
\end{itemize}

Pored konferencija, akademski časopisi igraju ključnu ulogu u diseminaciji AI istraživanja. Objavljivanje radova u visoko rangiranim časopisima omogućava dugoročnu dostupnost i vidljivost naučnih otkrića, što istraživačima pruža prepoznatljivost u globalnoj zajednici. Najpoznatiji časopisi u oblasti AI uključuju: Nature, Science, Journal of Artificial Intelligence Research (JAIR), IEEE Transactions on Pattern Analysis and Machine Intelligence, te Artificial Intelligence Journal.

Posljednjih godina, interes za AI konferencije značajno je porastao, prateći ubrzani razvoj i sve širu primjenu vještačke inteligencije u raznim sektorima. Stručnjaci, istraživači i profesionalci iz industrije okupljaju se na ovim događajima kako bi razmijenili najnovija saznanja, tehnologije i inovacije. Povećanje broja učesnika i raznovrsnosti konferencija odražava rastuću potrebu za istraživanjem u oblastima kao što su mašinsko učenje, robotika, obrada prirodnog jezika i autonomni sistemi. Vodeće konferencije sada primaju i do pet puta više radova nego prije deset godina, dok stopa prihvatanja radova ostaje stabilna oko 20\%.

\subsection{Digitalizacija kao preduslov za primjenu AI }

Od pisanja prethodnog rada (Ajanović et al. 2022), u posljednje dvije godine digitalizacija u Bosni i Hercegovini zabilježila je napredak u nekoliko ključnih oblasti, iako se suočava sa značajnim izazovima i zaostaje za susjednim državama. Pandemija COVID-19 ubrzala je ovaj proces, posebno u zdravstvu i obrazovanju, gdje su se digitalna rješenja morala brzo razvijati i implementirati. Infrastruktura se postepeno unapređuje kroz projekte poput IDDEEA/CIPS digitalnog potpisa, koji omogućava brže i sigurnije digitalno potpisivanje dokumenata (IDDEEA 2024). Također, pristupanje BiH programu Digital Europe stvara dodatne mogućnosti za razvoj digitalnih kapaciteta.

Iako postoje određeni napreci, Bosna i Hercegovina i dalje značajno zaostaje za susjednim državama, poput Hrvatske i Srbije, koje su već napravile značajne iskorake u digitalizaciji. U Hrvatskoj je uveden sistem e-Građani, koji omogućava građanima jednostavan pristup velikom broju javnih usluga putem interneta, a Srbija je razvila e-Zdravlje i e-Upravu, gdje građani mogu digitalno upravljati zdravstvenim kartonima i koristiti širok spektar usluga javne uprave online. Ovi napredni digitalni sistemi ne samo da ubrzavaju administrativne procese, već postavljaju temelj za primjenu tehnologija kao poput AI.

Nekoliko digitalnih rješenja u BiH je u fazi razvoja ili su već primijenjena. Adaptivno upravljanje saobraćajem i digitalno plaćanje karata u Sarajevu su u izradi, dok je zdravstveni sistem Kantona Sarajevo već u velikoj mjeri digitaliziran. Katastar u Federaciji BiH  je također digitalizovan, omogućavajući brži pristup informacijama o nekretninama, a uvođenje e-dnevnika u Kantonu Sarajevo omogućava praćenje ocjena i svjedočanstava. Digitalizacija biračkog procesa je u fazi pilot studije (IDDEEA 2024). Razvoj super-računara na Univerzitetu u Sarajevu, koji je još u početnoj fazi, mogao bi značajno unaprijediti istraživačke kapacitete, no ovi uspjesi su trenutno ograničeni na Kanton Sarajevo.

U oblasti digitalnih komunikacija, BiH i dalje zaostaje za susjednim zemljama. Prosječna brzina interneta u BiH iznosi 40,4 Mbps, dok Srbija i Crna Gora imaju brže mreže, a Sjeverna Makedonija dostiže gotovo 98 Mbps. Prema ICT Development Indexu Međunarodne telekomunikacione unije (ITU), BiH se nalazi na 80. mjestu od 175 zemalja, sa rezultatom 5.25 od 10, dok je u Evropi na pretposljednjem, 49. mjestu od 50 zemalja. Ovo jasno pokazuje potrebu za daljnjim ulaganjima u digitalnu infrastrukturu i razvoj informacijskih tehnologija (International Telecommunication Union 2023).

AI se uveliko oslanja na digitalne podatke i mreže za analizu velikih količina informacija, donošenje odluka u realnom vremenu i automatiziranja složenih procesa. Bez razvijene digitalne infrastrukture i uspostavljenih sistema, razvoj naprednih tehnologija poput AI je otežan i njihova efikasnost ograničena.

\subsection{Trendovi razvoja AI tehnologije}

U pogledu istraživanja, umjetna inteligencija je doživjela značajan iskorak naprijed. Dostignuća u velikim jezičkim modelima kao što su razne varijante GPT-4 modela od strane OpenAI (OpenAI et al. 2024), Claude modela iz Anthropic (Anthropic 2024), Gemini modela iz Google (Gemini Team et al. 2024) samo su neki od primjera da će se rapidni progres razvoja ovih tehnologija nastaviti i u budućnosti. Pored takmičenja u postavljanju novih rekorda na popularnim testnim problemima (Chiang et al. 2024), (Wang et al. 2024), kompanije su se aktivno posvetile problemu redukcije troškova korištenja ovih ogromnih modela. Raznim metodama kompresije kao što su kvantizacija parametara (Gholami et al. 2022) i destilacija znanja iz velikih “učitelj” modela u male “student” modele (Hinton, Vinyals, and Dean 2015), kompanije aktivno rade na smanjivanju broja parametara i preciznosti ovih modela kako bi smanjili potrošnju električne energije, te broj grafičkih kartica potrebnih za njihovo korištenje, i kako bi poboljšali iskustvo korisnika sa kraćim vremenom odziva cjelokupnog sistema. Neki od popularnih primjera su GPT-4o (OpenAI et al. 2024) i Gemini-Flash (Gemini Team et al. 2024), koji i pored velikog stepena kompresije pokazuju impresivne rezultate na popularnim testnim problemima.

Ujedno sa prethodnim dostignućima, oblast koja je doživjela znatan iskorak naprijed je i usklađivanje (alignment) modela sa ljudskim vrijednostima i namjerama. Ovaj koncept ima za cilj osigurati da modeli ostanu korisni i djeluju u skladu sa ljudskom etikom i ciljevima kako postaju autonomniji. U skladu s tim veliki broj tehnika poput RLHF (Ouyang et al. 2022), PPO (Schulman et al. 2017), DPO (Rafailov et al. 2023), i sličnih varijanti je doživio znatnu ekspanziju i prihvatanje od strane istraživača i inženjera u ovoj oblasti.

Važno je spomenuti da sva dostignuća iza zatvorenih vrata u prethodno spomenutim kompanijama su u korak praćena sa jednako velikim i značajnim postignućima u oblasti open-source modela. Pa tako, modeli poput Llama (Dubey et al. 2024), Mistral (Jiang et al. 2023), Qwen (Yang et al. 2024), DeepSeek (DeepSeek-AI et al. 2024) i mnogih drugih su uspjeli postići nivo kompetentnosti i iskoristivosti sa closed-source modelima poput GPT-4, Claude i Gemini. Zahvaljujući tome, krajnji korisnici danas imaju veliku mogućnost izbora između raznih modela koje mogu adaptirati za vlastite potrebe bez straha od dijeljenja privatnih podataka sa drugim kompanijama.

Razvoj jezičkih modela (LLMs) podjednako je praćen i razvojem vizuelno-jezičkih modela (VLMs) (Laurençon et al. 2024) koji pored tekstualnog ulaza od korisnika prihvataju i slike. Kao rezultat ove kombinacije, krajnjim korisnicima je otključan novi domen aplikacija u kojima se slike mogu koristiti kao mnogo bogatiji prikaz okoline u poređenju na tekstualni ulaz.  
Trenutni trendovi u primjeni jezičkih modela u robotici usmjereni su na rješavanje problema koji zahtijevaju više koraka kroz integraciju prirodnog jezika i akcijskih naredbi (Ahn et al. 2022). Jezički modeli, omogućavaju robotima da bolje razumiju i interpretiraju složene upute, prevodeći ih u akcije (Luijkx et al. 2022). U kombinaciji sa metodama planiranja, ova tehnologija omogućava robotima da izvršavaju zadatke koji zahtijevaju planiranje, prilagođavanje novim informacijama i donošenje odluka u stvarnom vremenu. Takođe, primjena ovih modela olaksava robotima da iskoriste “zdravorazumsko znanje” i brzo prilagođavaju raznim svakodnevnim kontekstima (Ma et al. 2024).

Moderni računarski sistemi, uključujući one koji koriste napredne AI modele poput ChatGPT-a, troše ogromne količine energije i postaju neodrživi za dugoročni razvoj još složenijih modela. Iako su specijalizirani čipovi poput GPU-a i TPU-a omogućili ogroman napredak u performansama AI sistema, njihov energetski otisak (u velikim data centrima) predstavlja sve veći problem. Svaki dalji razvoj složenih AI modela zahtijeva značajne resurse, što dovodi do povećanja energetske potrošnje data centara, koji trenutno troše oko 200 teravat-sati energije godišnje, a očekuje se da će ta potrošnja drastično rasti do 2030. godine. Dodatno, aplikacije poput Interneta stvari (IoT) i autonomnih robota, koje možda ne koriste uvijek intenzivne algoritme AI-a, također moraju smanjiti energetsku potrošnju kako bi postale održive. S obzirom na to da se potražnja za računarskom snagom udvostručuje svaka dva mjeseca (Mehonic and Kenyon 2022), postaje jasno da trenutni model razvoja AI tehnologija nije održiv na duže staze.

Dalji napredak umjetne inteligencije ovisit će o razvoju novih računarskih tehnologija, počevši od najnižeg nivoa, tj. razvojem novih elektronskih komponenti poput kompjuterske memorije (Mehonic et al. 2020). Zatim će biti potreban razvoj još specijaliziranijih čipova, a na kraju će se istraživati novi koncepti, kao što su neuromorfološka (Mehonic and Eshraghian 2023) ili druge nekonvencionalne metode obrade podataka i signala. Ovi novi pristupi mogli bi značajno smanjiti energetsku potrošnju, omogućiti veću računarsku moć te održiv razvoj AI sistema, čime bi se riješili rastući problemi energetske potrošnje.

\subsection{Etika, Regulacija, Standardi i Javna Uprava}

Negativni rizici AI još uvijek nisu dovoljno istraženi niti jasno definisani, a trenutno ne postoje precizne metode za njihovo mjerenje (NIST 2023). Ipak, pojavili su se pokušaji procjene i upravljanja ovim rizicima kroz prvu svjetsku regulativu o AI, Evropski akt o umjetnoj inteligenciji – EU AI Act (EU 2024), kao i kroz nekoliko međunarodnih standarda i tehničkih izvještaja koji se bave upravljanjem i sigurnošću AI (npr. ISO/IEC 42001:2023, ISO/IEC 23894:2023, ISO/IEC TR 5469:2024). Također, u SAD-u, Kalifornija je uvela svoj vlastiti zakon o AI pod nazivom SB 1047, koji ima za cilj regulisanje upotrebe AI tehnologija na državnom nivou, s fokusom na transparentnost i odgovornost.

AI sistemi predstavljaju jedinstvene izazove u poređenju s tradicionalnim softverskim inženjeringom, posebno zbog svoje stohastičke prirode i sposobnosti autonomnog učenja iz podataka. Ovi izazovi uključuju nedostatak transparentnosti i objašnjivosti – često nazivani problemom "crne kutije", gdje su unutrašnji procesi AI modela nejasni ili teško razumljivi čak i onima koji su ih razvili. Dodatno, AI sistemi ponekad proizvode halucinacije – slučajeve kada generativni modeli kreiraju informacije koje nisu tačne ili zasnovane na podacima, što može dovesti do ozbiljnih grešaka u realnim aplikacijama.

Primjeri iz cijelog svijeta pokazuju kako je upotreba AI bez rješavanja ključnih rizika već prouzrokovala štetu, uključujući diskriminaciju pri zapošljavanju, probleme u provođenju zakona i incidente u sigurnosno-kritičnim operacijama.

Potreba za regulacijom razvoja, primjene i korištenja AI kroz standarde, regulative i politike se posebno povećala nakon široke upotrebe generativne AI od novembra 2022. godine. EU AI Act, koji je stupio na snagu u augustu 2024., uvodi zakonske obaveze u skladu s nivoom rizika koji AI sistemi predstavljaju za ljude, društvo i okoliš. AI sistemi su prema ovom aktu podijeljeni u četiri kategorije rizika: neprihvatljiv, visok, ograničen i nizak, uz dodatnu kategoriju za AI za generalnu upotrebu (vidi Tabelu 1).

\begin{itemize}
    \item \textbf{Neprihvatljivi AI sistemi}: poput onih koji koriste biometrijsku identifikaciju u realnom vremenu u javnim prostorima ili manipulišu ponašanjem korisnika, biće zabranjeni.
    \item \textbf{Visokorizični sistemi}: uključujući one koji se koriste u zdravstvu, obrazovanju, zapošljavanju i kritičnoj infrastrukturi, moraće ispunjavati stroge provjere usklađenosti s regulativama koje osiguravaju sigurnost i zaštitu osnovnih prava korisnika.
    \item \textbf{Sistemi s ograničenim rizikom}: poput onih koji omogućuju generisanje ili manipulaciju slikama, zvukom ili videom (npr. deepfakes), zahtijevaju mjere transparentnosti kako bi korisnici znali da komuniciraju s AI.
    \item \textbf{Niskorizični sistemi}: kao što su filteri za neželjenu poštu, nemaju posebne regulatorne zahtjeve i mogu se koristiti uz opći kodeks ponašanja.
    \item \textbf{AI za generalnu upotrebu}: poput fondacijskih modela kao što su ChatGPT i drugi modeli opšte namjene, podliježu posebnim zahtjevima za transparentnost. Ako se koriste u visokorizičnim situacijama ili imaju značajan uticaj na ljude, moraju proći dodatne evaluacije.
\end{itemize}

\begin{table}[h!]
\centering
\caption{EU AI Act: Obaveze i vremenski okvir za provedbu zakona po kategoriji rizika}
\begin{tabular}{|l|l|l|}
\hline
\rowcolor{gray!30} \textbf{Kategorija rizika} & \textbf{Zakonska obaveza} & \textbf{Vremenski okvir za primjenu} \\ \hline
\rowcolor{red!20} Neprihvatljiv & Zabranjeni & Februar 2025 (6 mjeseci) \\ \hline
\rowcolor{orange!20} Visok & Provjere usklađenosti & August 2027 (36 mjeseci) \\ \hline
\rowcolor{yellow!20} Ograničen & Mjere transparentnosti & August 2025 (12 mjeseci) \\ \hline
\rowcolor{green!20} Nizak & Opći kodeks ponašanja & Nije primjenjivo \\ \hline
\rowcolor{gray!10} AI za generalnu upotrebu & Transparentnost i evaluacija & August 2025 (12 mjeseci) \\ \hline
\end{tabular}
\end{table}

Osim Evropskog zakona o AI, Kalifornija je u avgustu 2024. usvojila SB 1047 (Davis 2024), zakon koji postavlja okvir za regulaciju naprednih AI sistema s naglaskom na sigurnost, transparentnost i odgovornost. Zakon se fokusira na "frontier" AI modele, odnosno napredne sisteme koji zahtijevaju značajne resurse za obuku i imaju potencijalno visoke rizike. Kompanije koje razvijaju ove modele moraju uspostaviti sigurnosne protokole prije obuke, uključujući mogućnost hitnog isključivanja u slučaju problema. Uz to, zakon nalaže redovne revizije i obavezu prijavljivanja incidenata u roku od 72 sata.

SB 1047 zabranjuje upotrebu AI modela ako postoji ozbiljan rizik da bi mogli prouzrokovati štetu, poput cyber napada ili razvoja oružja. Zakon uključuje i osiguranu pravnu zaštitu zaposlenicima u slučaju prijavljivanja nepravilnosti. Pored ovih mjera, zakon uspostavlja CalCompute, državni računski sistem koji pruža podršku akademskim institucijama i startupima u razvoju AI tehnologija.

Ovaj zakon je prvi te vrste u SAD-u i predstavlja značajan korak ka regulaciji AI sistema na državnom nivou, uz kombinaciju sa EU AI Act. Ovi zakoni imaju za cilj osigurati etičan i odgovoran razvoj AI tehnologija, uz visok nivo zaštite ljudskih prava i sigurnosti.

\section{Organizacija naučno-istraživačkog djelovanja}

Ovo poglavlje se bavi ključnim aspektima koji čine temelj uspješnog naučno-istraživačkog rada, sa fokusom na razlikovanje između istraživanja i edukacije, te stvaranje preduslova za dugoročni razvoj istraživačkih kapaciteta. Razmatraju se osnovni principi istraživačkog rada, uključujući ulogu institucija koje predvode inovacije i globalni napredak u umjetnoj inteligenciji. Kroz primjere istaknutih centara kao što su Bell Labs, Google DeepMind, Institut za vještačku inteligenciju Srbije te Komet centri i Institut za nauku i tehnologiju Austrije u Austriji, ilustriraju se različiti modeli i pristupi razvoju istraživačke izvrsnosti.

\subsection{Istraživanje i edukacija}

Istraživanje i edukacija predstavljaju dva ključna aspekta akademskog i stručnog razvoja, a iako se često isprepliću, imaju različite ciljeve. Istraživanje se fokusira na stvaranje novih saznanja i pomicanje granica onoga što znamo o svijetu. Kroz istraživanje dolazimo do novih otkrića, rješenja za kompleksne probleme i inovacija koje unapređuju nauku i tehnologiju. U tom kontekstu, istraživači su usmjereni na produbljivanje razumijevanja i otkrivanje novih informacija koje mogu koristiti društvu u cjelini.

S druge strane, edukacija je usmjerena na prenošenje postojećih znanja i vještina na studente, profesionalce i širu javnost. Njena svrha je osposobiti pojedince da primijene ta znanja u praksi, bilo kroz rad ili daljnje istraživanje. Edukacija postavlja temelje za kritičko mišljenje i razvoj vještina, čime se stvara nova generacija istraživača i stručnjaka.

Iako su po svojoj prirodi različiti, istraživanje i edukacija se međusobno nadopunjuju. Rezultati istraživanja obogaćuju obrazovne programe, dok edukacija pruža okvir u kojem se razvijaju budući istraživači. Sinergija između ova dva procesa ključna je za kontinuirani napredak i inovacije u svim područjima znanja. Samo kombinacijom kvalitetne edukacije i snažnog istraživačkog kadra moguće je postići održiv razvoj industrije i osigurati konkurentnost na globalnom tržištu. Dok edukacija omogućava širenje znanja i upotrebu postojećih tehnologija, istraživanje stvara nove prilike za napredak i inovacije, što je ključno za dugoročni uspjeh ne samo u vještačkoj inteligenciji, već i u industriji općenito. Omjer naučno-istraživačkog rada i edukacije dodatno kategoriše akademske ustanove na univerzitete, institute, više škole, itd. 

\subsection{Preduslovi za uspješan naučno-istraživački rad}

Uspješan naučno-istraživački rad ne proizlazi samo iz puke primjene pojedinačnih faktora, već je rezultat harmonične integracije različitih preduslova koji zajedno osiguravaju kvalitet, efikasnost i održivost istraživanja. Na prvom mjestu je \textit{jasna misija i vizija} koja je u skladu s naučnim ciljevima. Ovo daje smijer i svrhu istraživačkih aktivnosti i pomaže u postavljanju prioriteta. \textit{Stručno osoblje}. Organizacija treba da privuče i zadrži talentovane naučnike, istraživače i tehničko osoblje, neophodno za provođenje visokokvalitetnog istraživanja. \textit{Jako vodstvo i menadžment} je ključno za vođenje naučnog programa organizacije, obezbjeđivanje finansiranja i podsticanje produktivnog istraživačkog okruženja jer osigurava da se resursi efikasno raspoređuju i da su istraživačke aktivnosti dobro koordinisane. \textit{Strateško planiranje}, tj. njihov razvoj i implementacija koji ocrtavaju istraživačke ciljeve, prioritete i prekretnice pomaže u usklađivanju napora i praćenju napretka ka postizanju naučnih ciljeva. \textit{Adekvatna finansijska sredstva i resursi} za podršku istraživačkim aktivnostima koje uključuje finansiranje istraživačkih projekata, opreme, objekata i pomoćnog osoblja. \textit{Dobra istraživačka infrastruktura}, tj. pristup naprednoj laboratorijskoj opremi, tehnologiji i sistemima za upravljanje podacima kao i dobro održavana i ažurna infrastruktura za testiranje i prikupljanje podataka su neophodni za rad. \textit{Kontrola i evaluacija kvaliteta} kroz implementaciju robusnih mjera kontrole kvaliteta i redovno i kontinuirano vrednovanje istraživačkih procesa i ishoda pomažu da se osigura da je naučni rad visokog kvaliteta i da ispunjava utvrđene standarde. \textit{Inovativnost i fleksibilnost} kroz poticanje inovacija i prilagodljivost novim metodama, tehnologijama i promjenama u naučnom okruženju može pomoći organizaciji da ostane na čelu istraživanja. \textit{Poštovanje etičkih standarda i regulatornih zahtjeva} u istraživanju je ključno za održavanje integriteta i kredibilitete. \textit{Kultura podrške} koja vrednuje radoznalost, uvažavanje, kritičko mišljenje i kontinuirano učenje njeguje okruženje pogodno za naučna otkrića i izvrsnost. \textit{Podsticanje unutarnje saradnje} i saradnje sa drugim istraživačkim institucijama, univerzitetima i industrijom, može poboljšati kvalitet, budžet i uticaj naučnog rada i naučne rezultate. \textit{Učinkovita komunikacija} kroz unutrašnje i vanjske kanale komunikacije (npr. naučne radove, medije, društvene mreže, itd.) važni su za dijeljenje rezultata istraživanja i promoviranje rada organizacije.

\subsection{Osnovne istraživačkog rada}

Istraživački rad u akademskim ustanovama temelji se na nekoliko principa i struktura koje osiguravaju efikasnost i kvalitet istraživanja. Cilj ove sekcije je uvesti koncepte i terminologiju. Osnovna ćelija naučno-istraživačkog rada jeste \textit{istraživačka grupa}, koja okuplja istraživače s različitim znanjima i vještinama ali je obično usko fokusirana na jednu temu ili oblast. \textit{Principal Investigator (PI)} vodi projekt te je uključen u sve aspekte rada, od početne ideje do implementacije, publikacije i prezentacije. Ostali članovi grupe su postdoktorski istraživači te doktorski studenti. \textit{Doktorski studenti} se najčešće specijalizuju u uskoj tematici te doprinose njenom proširenju što čini osnovu doktorske disertacije. \textit{Postdoktorski istraživači} svojim tehničkim znanjem i zalaganjem doprinose rješavanju ključnih problema. Tehničko osoblje je neophodno kako bi se osigurao nesmetan rad laboratorija. Pored toga, neke grupe uključuju master i bachelor studente u svoj rad. Doprinosi uspješne grupe rezultiraju značajnim napretkom u naučnom polju, kroz nove teorije, metodologije i primjene. 

Istraživačka ustanova se sastoji iz više pojedinačnih istraživačkih grupa koje su dalje organizovane u odsjeke grupisane po tematici istraživanja. \textit{Vertikalna} istraživanja fokusiraju se na dubinsko proučavanje specifičnih tema, dok \textit{horizontalna} istraživanja obuhvataju širi spektar tema unutar jednog ili više polja. Horizontalna istraživanja često zahtijevaju kolaboraciju više istraživačkih grupa unutar jedne ili više akademskih ustanova. Ova dva pristupa se međusobno nadopunjuju, omogućavajući cjelovito razumijevanje kompleksnih problema. Istraživanja mogu biti \textit{aplikativna}, usmjerena na direktnu primjenu rezultata za rješavanje praktičnih problema, ili \textit{fundamentalna}, fokusirana na sticanje osnovnog znanja i razumijevanja bez neposredne praktične primjene. 

Istraživačke zajednice djeluju na lokalnom i globalnom nivou. \textit{Lokalna} istraživanja često se fokusiraju na specifične probleme i resurse dostupne unutar određene regije, dok \textit{globalne} zajednice omogućavaju širu razmjenu ideja i resursa. Naučno-istraživačke konferencije ključne su za povezivanje ovih zajednica i promovisanje kolaboracije.

Interni seminari, kolokvijumi i drugi oblici stručnih sastanaka unutar istraživačke institucije igraju ključnu ulogu u unapređenju naučno-istraživačkog rada. Ovi događaji omogućavaju istraživačima da prezentiraju svoje radove, dobiju povratne informacije i razvijaju nove ideje kroz interakciju s kolegama. Istraživačke institucije mogu imati različit stepen specijalizacije. Specijalizirane institucije fokusiraju se na uska područja istraživanja, omogućavajući duboku ekspertizu, dok institucije sa širokim spektrom istraživanja obuhvataju različite discipline, potičući interdisciplinarne projekte i inovacije.

Efikasno upravljanje istraživačkom grupom zahtijeva administrativnu podršku, financijske resurse i pristup potrebnim alatima i tehnologijama. Podrška istraživačkim grupama kroz \textit{projekte} i \textit{grantove} ključna je za poticanje inovacija i napretka. Finansiranje istraživačkih projekata dolazi iz različitih izvora, uključujući državne grantove, industrijska partnerstva i međunarodne fondove. Prijedlozi za projekte mogu biti \textit{bottom-up}, gdje istraživači sami iniciraju projekte, ili \textit{top-down}, gdje organizacije postavljaju istraživačke prioritete.

\subsection{Primjeri institucija}

Kao konkretne primjere organizacije naučno-istraživačkog djelovanja predstavljamo nekoliko istaknutih institucija koje su postigle značajan utjecaj na globalnom nivou, ugled u naučnim krugovima ili su potekle iz sličnog okruženja kao u Bosni i Hercegovini.

\subsubsection{Bell Labs}
Bell Labs, službeno poznat kao Bell Telephone Laboratories, Inc., osnovan je 1925. godine kao istraživački i razvojni ogranak kompanije American Telephone and Telegraph Company (AT\&T). Glavni cilj osnivanja Bell Labs-a bio je unaprijediti i održavati tehnološku superiornost Bell System-a, mreže kompanija pod kontrolom AT\&T-a, odgovornih za cjelokupnu telekomunikacijsku infrastrukturu SAD-a.

Bell Labs je bio finansiran od strane AT\&T-a, koji je ostvarivao značajan profit zahvaljujući svom telekomunikacijskom monopolu. Finansijska stabilnost omogućila je Bell Labs-u da provodi dugoročna istraživanja, ulažući značajna sredstva u fundamentalnu nauku, bez pritiska za kratkoročnu dobit. Kontinuirano finansiranje osiguralo je potrebnu stabilnost za dugoročni razvoj fundamentalne i primijenjene nauke (Gertner 2012).

Iako je Bell Labs imao širok obim istraživanja, glavni fokus bio je na kreiranju globalno povezane telekomunikacijske mreže. Laboratorija je ostvarila ključne doprinose u oblastima kao što su fizika čvrstog stanja, teorija informacija i telekomunikacije, te u razvoju revolucionarnih tehnologija poput tranzistora, lasera, satelita, C programskog jezika i UNIX operativnog sistema.

Organizacijska struktura Bell Labs-a bila je dizajnirana s naglaskom na saradnju između različitih disciplina. Mervin Kelly, predsjednik Bell Labs-a tokom “zlatnog doba inovacija”, naglašavao je važnost prostora koji olakšava komunikaciju među naučnicima i okruženja koje podstiče slobodnu razmjenu ideja. Umjesto stroge podjele rada po odjelima, naučnici su imali slobodu da istražuju i sarađuju, što je uzrokovala mnogim od pomenutih inovacija.

Eksterna saradnja bila je još jedan ključni aspekt rada Bell Labs-a. Laboratorija je održavala snažne veze s univerzitetima, vladinim agencijama i drugim industrijskim istraživačkim centrima. Saradnja je često uključivala zajedničke projekte koji su koristili resurse i stručnost Bell Labs-a. Na primjer, tokom Drugog svjetskog rata, Bell Labs je sarađivao s vladom SAD-a na razvoju radara, naprednim komunikacijskim sistemima i kriptografiji, što su bile ključne tehnologije za pobjedu saveznika.

Uspjeh Bell Labs-a mjerio se kroz doprinos tehnološkom liderstvu AT\&T-a i kroz uticaj na šire društvo. Laboratorija je proizvela zapanjujući broj inovacija, uključujući deset Nobelovih nagrada, a mnogi uposlenici su nastavili sa značajnim doprinosima u nauci i tehnologiji, te postali ključni igrači u formiranju nove struje inovacija – Silikonske doline.

\subsubsection{Google DeepMind}

Google DeepMind, osnovan 2010. godine u Ujedinjenom Kraljevstvu kao DeepMind Technologies, je britansko-američka istraživačka laboratorija za umjetnu inteligenciju. Osnivači Demis Hassabis, Shane Legg i Mustafa Suleyman pokrenuli su kompaniju s misijom "rješavanja inteligencije s ciljem unaprjeđenja čovječanstva". Početne investicije od strane tehnoloških lidera poput Elona Muska i Petera Thiela, omogućile su fokus na fundamentalna istraživanja u oblasti AI. Google je 2014. godine kupio DeepMind za oko 500 miliona funti (oko 1.6 milijardi KM u današnjoj vrijednosti), osiguravajući resurse za dugoročni razvoj istraživanja.

Od početka, DeepMind je kombinovao osnovna istraživanja s razvojem tehnologija s praktičnim utjecajem. Inspirisan organizacijom Bell Labs-a, DeepMind promoviše interdisciplinarnu saradnju i slobodnu razmjenu ideja. U ranim danima, organizacija je bila podijeljena na DeepMind Research, fokusiran na fundamentalna istraživanja, i DeepMind Applied, fokusiran na primijenjena istraživanja. Iako se struktura kroz godine mijenjala, interakcija fundamentalnog i primijenjenog istraživanja ostala je ključna.

Prvi veliki uspjeh DeepMind-a bio je razvoj AI algoritma koji je 2013. godine postigao nivo ljudskih sposobnosti u igranju Atari igara, demonstrirajući sposobnost učenja bez ljudske intervencije (Mnih et al. 2015). Ovaj uspjeh privukao je pažnju Google-a i bio katalizator za akviziciju godinu poslije. Kasniji projekti, poput AlphaGo i AlphaZero, dodatno su popularizirali AI i prikazali potencijal novog vala umjetne inteligencije u kompleksnim igrama poput šaha i kineskog Go-a (Silver et al. 2016).

Među najznačajnijim dostignućima DeepMind-a je AlphaFold, algoritam koji je riješio dugogodišnji problem predviđanja strukture proteina, jednog od fundamentalnih problema u biologiji (Jumper et al. 2021). AlphaFold je omogućio kreiranje baze struktura gotovo svih poznatih proteina (preko 200 miliona), čime je postavio novu osnovu za biomedicinska istraživanja i razvoj novih lijekova. DeepMind je također pionir u razvoju transformera (Vaswani et al. 2023), ključne tehnologije za razvoj modernih jezičkih modela i chatbotova. Ovi modeli, poput ChatGPT-a i Gemini, značajno su unaprijedili sposobnost AI-ja da razumije i generira ljudski jezik, što je dovelo do moderne AI revolucije koju vidimo danas.

Osim tehnoloških inovacija, DeepMind se bavi i etičkim pitanjima u razvoju AI-ja. Kompanija je osnovala jedinicu DeepMind Ethics \& Society i formirala etički odbor kako bi adresirala pitanja poput pristranosti, transparentnosti i odgovorne upotrebe tehnologije. Kroz saradnju s univerzitetima, istraživačkim institutima i industrijskim partnerima, DeepMind nastoji proširiti utjecaj svojih istraživanja i AI tehnologija na globalnom nivou.

\subsubsection{Institut za vještačku inteligenciju Srbije}

Istraživačko-razvojni institut za vještačku inteligenciju Srbije (IVI) osnovan je 2021. godine u Novom Sadu od strane Vlade Republike Srbije, uz kontinuiranu podršku kako kroz finansiranje, tako i kroz prilagođavanje zakonodavnog okvira (Vlada Republike Srbije 2021). Vlada Srbije osigurala je početna sredstva za rad Instituta, izdvojivši 200.000.000 dinara (oko 3.5 miliona KM), čime je omogućila stabilan početak ovog ambicioznog projekta. Institut danas zapošljava ukupno 59 radnika, od čega je 51 istraživač, dok 8 zaposlenih radi u administraciji.

Od osnivanja, IVI je bio ključan partner u sprovođenju nacionalne strategije razvoja vještačke inteligencije i aktivno učestvuje u međunarodnim projektima. Trenutno učestvuje u 4 Horizon Europe projekta, 4 međunarodna projekta (uključujući Interreg program), 2 nacionalna projekta podržana od Fonda za nauku Srbije, te sprovodi 4 projekta sa javnim institucijama (IVI 2024). Pored toga, Institut je podnio 4 patentne prijave. IVI također radi na 13 komercijalnih projekata, koji uključuju saradnju s industrijskim partnerima iz zemlje i inostranstva.

Primjer uspješnog projekta je saradnja sa Telekom Srbija, gdje je IVI implementirao napredne AI tehnike za analizu korisničkih emailova, koristeći metode modeliranja tema za automatsko sortiranje i odgovaranje na korisničke zahtjeve, čime je značajno unaprijeđena efikasnost korisničke podrške.

Institut aktivno doprinosi jačanju AI ekosistema u Srbiji i regionu, postavljajući temelje za buduće generacije istraživača i lidera. Kroz međunarodnu saradnju i projekte, IVI osigurava da Srbija bude konkurentna u globalnoj AI ekonomiji (Kisačanin, Vasiljević, and Ćulibrk 2021)

\subsubsection{Komet Centri u Austriji}

Komet Centri (K2) u Austriji predstavljaju napredan model za realizaciju strateških istraživačkih programa u oblasti veštačke inteligencije i srodnih tehnologija. Ovi centri organizovani su kao konzorcijumi koji kombinuju snage naučnih i komercijalnih partnera kako bi se ostvarili visokokvalitetni istraživački ciljevi (FFG 2024).

Komet Centri zahtijevaju barem jednog naučnog partnera i pet kompanijskih partnera, što omogućava kombinaciju akademskog i komercijalnog znanja i resursa. Programi su dizajnirani da traju osam godina, sa finansiranjem koje se kreće od 40\% do maksimalno 55\% iz javnih sredstava. Kompanijski partneri doprinose najmanje 40\%, dok naučni partneri obezbjeđuju minimalno 5\% finansiranja. Najveće federalno finansiranje iznosi 4 miliona eura godišnje, dok provincijsko finansiranje može doseći do 2 miliona eura godišnje.

Komet Centri su razvili uspješne primjere u istraživačkom sektoru, kao što su Austrian Blockchain Center (ABC Research GmbH), Know-Center GmbH, Linz Center of Mechatronics GmbH, i Virtual Vehicle Research GmbH. Svaka od ovih institucija doprinosi inovacijama u svojim specifičnim oblastima: Austrian Blockchain Center (ABC Research GmbH) fokusira se na napredne istraživačke projekte u oblasti blockchain tehnologije, sa ciljem unapređenja sigurnosti i efikasnosti u digitalnim transakcijama. Know-Center GmbH je istraživački centar specijalizovan za analizu podataka i big data, koji se bavi razvojem novih tehnika za ekstrakciju vrednih uvida iz velikih količina podataka. Linz Center of Mechatronics GmbH kombinira mehatroniku i robotiku, pružajući rješenja za automatsku kontrolu i pametne sisteme koji se primjenjuju u industriji i proizvodnji. Virtual Vehicle Research GmbH se fokusira na virtualno testiranje i simulaciju za automobilski sektor, pomažući u razvoju i optimizaciji vozila kroz napredne simulacijske tehnologije.

Kroz ovakve center, Austrija ne samo da podržava napredak u ključnim tehnološkim oblastima, već i jača svoj položaj kao lider u globalnoj istraživačkoj zajednici, pružajući inovativna rješenja i stvarajući mogućnosti za buduće generacije istraživača i preduzetnika.

\subsubsection{Institut za nauku i tehnologiju Austrije}

Institut za nauku i tehnologiju Austrije (ISTA) je veoma mladi međunarodni institut posvećen postdiplomskom obrazovanju i istraživanjima u oblastima prirodnih, matematičkih i računarskih nauka. ISTA-u je osnovala Savezna vlada Austrije i Vlada Donje Austrije 2006. godine. Kampus je otvoren 2009. godine u gradu Klosterneuburgu, na periferiji Beča. Institut je osnovan na osnovu skupa principa koje su prvi formulirali Haim Harari, Olaf Kübler i Hubert Markl (Harari, Kuebler, and Markl 2006). 

Uz javna i privatna sredstva koja su im povjerena, institut ima za cilj dostići svoj puni kapacitet od 150 istraživačkih grupa do 2036. godine. Pored fokusa na istraživanje, ISTA je razvila program podrške za komercijalizaciju naučnog istraživanja kroz xista program. Pored novčane pomoći, kroz otvaranje xista inovacijskog centra pri institutu istraživačima se obezbjeđuje punu podrška tokom cijelog procesa transformiranja naučnih otkrića u industrijske i komercijalne aplikacije. 

Svjedočanstvo njihovog dobrog rada prepoznato je kroz dva glavna dostignuća: kada se normalizira prema veličini, istraživački rezultat instituta je rangiran kao treći u svijetu prema Nature Index-u u 2019. godini, dok je ISTA centar prikupio 45 miliona eura u prvoj iteraciji finansiranja istraživačkih spin-off kompanija u 2021. godini (ISTA 2019).

\section{Preporuka za BiH}

Ovo poglavlje nudi preporuke za unapređenje naučno-istraživačkog rada u Bosni i Hercegovini, uključujući osnivanje javnog AI instituta, kontinuiranu eksterne evaluaciju istraživača, podršku karijerama doktoranata u industriji, uspostavljanje višegodišnjeg fonda za nauku i reformu zakona o javnim nabavkama. Također se predlaže usklađivanje s EU standardima i rad na strategiji za dugoročni razvoj AI.

\subsection{Interdisciplinarni, javni AI Institut}

Trenutno, Bosna i Hercegovina zaostaje u području naučnoistraživačkog rada, ne samo u odnosu na Europu, već i u odnosu na zemlje u regionu. U poređenju sa Hrvatskom, Slovenijom i Srbijom, BiH ima pet do deset puta manje istraživača na milion stanovnika (Our World in Data 2024) te ulaže pet do deset puta manje svog BDP-a u istraživanje i razvoj (Our World in Data 2024). Manjak ulaganja se ogleda i u broju napisanih radova, gdje naši istraživači imaju napisano ukupno šest do sedam puta manje radova te su dva puta manje citirani (SJR 2024).  Jedan od načina na koji bi se takva razlika mogla smanjiti jeste osnivanjem institucija koje bi se fokusirale prvenstveno na naučnoistraživački rad, poput instituta. 

S obzirom na multidisciplinarnost i široku primjenjivost vještačke inteligencije, osnivanje instituta fokusiranog na vještačku inteligenciju moglo bi biti idealno rješenje za unapređenje naučnih rezultata u mnogim oblastima. Važno je naglasiti da cilj javno osnovanog instituta ne bi trebalo da bude isključivo dizajniranje proizvoda ili patenata (iako bi to moglo biti nuspojava), već novih naučnih spoznaja te izgradnja svjetski priznatih stručnjaka iz baze domaćih talenata, čiji bi rad bio međunarodno priznat i koji bi bili u toku s najnovijim dostignućima u polju vještačke inteligencije. Osim što bi našim vrhunskim talentima pružili priliku da se bave naukom u svojoj zemlji, umjesto u inostranstvu, njihovo zadržavanje u lokalnom ekosistemu omogućilo bi ostalim akterima, od studenata istraživača do privatnih kompanija, da imaju koristi od njihovog znanja. Uzimajući u obzir pozitivne efekte koje bi imali na stručnost istraživača i efikasnost domaćih kompanija, troškovi njihovog naučnoistraživačkog rada bi se vremenom višestruko isplatili.

Pored toga, zbog samog vala tehnoloških promjena koje potencijalno donosi AI (Alonso et al. 2022), veoma je bitno da se na javnom institutu izučavaju te teme i budu angažovani istraživači koji će brinuti u javnom interesu, rizicima koje AI donosi za društvo te dostupnosti benefita koje AI donosi široj zajednici i samoj suverenosti u doba AI (Mügge 2024).

\subsection{Kontinuirana eksterna evaluacija}

Uvođenje kontinuirane eksterne evaluacije istraživača i institucija ključno je za osiguranje objektivnosti i stalnog unapređenja kvaliteta naučnih doprinosa. NIR treba redovno evaluirati ne samo na osnovu broja i kvaliteta objavljenih radova prema međunarodnim rangiranjima poput Core Ranking-a ili h-indeksa, već i prema učešću u međunarodnim projektima, umreženosti s vodećim globalnim institucijama i uspješnosti kolaboracija. Kontinuirana evaluacija bi omogućila praćenje napretka kroz vrijeme, jačanje međunarodne kompetitivnosti i poticanje stalnog inoviranja i transfera znanja unutar naučne zajednice.

\subsection{Karijerne prilike za doktorante izvan akademskih institucija}

Iako je uobičajeno da se doktorati realizuju kroz zaposlenje na univerzitetima, postoji i alternativni pristup koji je usmjeren na istraživački rad u industriji i rješavanje praktičnih problema s kojima se kompanije suočavaju. Doktorati povezani s industrijom omogućavaju studentima da direktno rade na istraživačkim projektima unutar kompanija. Ove mogućnosti ne samo da otvaraju nova radna mjesta, već i pružaju doktorantima priliku da razviju inovativna rješenja za stvarne poslovne izazove, čime doprinose razvoju kompanija i napretku industrijskih sektora.

Uspostavljanje bolje saradnje između univerziteta i industrije može značajno poboljšati kapacitet za istraživanje i razvoj, te ojačati nacionalnu ekonomiju. Doktorati zasnovani na industrijskim potrebama mogu povezati visoko obrazovanje s poslovnim sektorom, stvarajući sinergiju između znanja i primjene. Ovaj pristup nudi pobjednički scenarij za obje strane – univerziteti dobivaju pristup realnim problemima i resursima, dok kompanije dobivaju visokoobrazovane stručnjake spremne za istraživanje i inovacije. Ovakav vid saradnje je neodrživ bez podrške države sa raznim poreskim olakšicama i subvencioniranjem troškova. Također, u zapadnim zemljama ustaljena je praksa da na mnogim odgovornim pozicijama u kompanijama budu angažovani stručnjaci sa doktorskom titulom.

\subsection{Fond za nauku i zakonski okvir}

Fond za nauku s višegodišnjim finansiranjem bi pružio stabilniju podršku za dugoročne istraživačke projekte. Trenutni sistem, baziran na jednogodišnjim projektima, ne podržava adekvatno dugoročno planiranje i realizaciju kompleksnih istraživačkih aktivnosti. Višegodišnje finansiranje omogućilo bi istraživačima da se fokusiraju na projekte od strateškog značaja i olakšalo usklađivanje sa evropskim standardima finansiranja, kao što su oni u okviru programa Horizon Europe. Fondovi u Srbiji (Republika Srbija 2018) i Hrvatskoj (HRZZ 2009) pokazali su se uspješnim u unapređenju istraživačkih kapaciteta i kvaliteta naučnog rada. Reforma zakona o javnim nabavkama je ključna za ubrzanje i efikasniju realizaciju naučnih istraživanja. Trenutni procesi nabavki su spori i birokratski, što otežava nabavku specijalizirane istraživačke opreme. 

\subsection{AI Etika, regulative i ministarstva}

Da bi se pridržavale strogih regulativa, zemlje članice EU će uskoro biti obavezne da uspostave određena nadležna tijela odgovorna za nadzor usklađenosti unutar svojih jurisdikcija. Očekuje se da će ova tijela osigurati prijavljivanje ozbiljnih incidenata vezanih za AI, olakšati registraciju sistema AI s visokim rizikom u nadolazećoj EU bazi podataka, te koordinirati s nedavno uspostavljenim EU Uredom za AI (EU AI Office). Pored toga, nekoliko članica je već uvelo zasebno Ministarstvo za digitalni rad i tehnologiju, te brojni dobrovoljni projekti su se pojavili širom Evrope i svijeta, podržavajući istraživanja u oblasti AI, zelenog softvera i održivog razvoja, s ciljem unapređenja nacionalne kompetencije i razvoja talenata u oblasti AI. 

\subsection{Strategija za razvoj AI}

Za kraj, kao i pri prethodnom radu AI vizije (Ajanović et al. 2022), i dalje smatramo da je prvenstveno potrebno razviti jasnu strategiju po uzoru na predstavljene primjere drugih zemalja koja bi omogućavala dugogodišnji razvoj AI u Bosni i Hercegovini. Sam proces izrade strategije, kao dio od krucijalnog značaja, treba biti inkluzivan. Tokom prikupljanja informacija i mišljenja i same izrade strategije potrebno je uključiti sve relevantne aktere, uključujući akademsku zajednicu, javnu upravu, predstavnike industrije i civilnog društva kao i neovisne strane eksperte. Posebnu pažnju bi trebalo posvetiti brendiranju strategije i njene vidljivosti sa ciljem kreiranja pozitivnog ugleda u Bosni i Hercegovini kao i inostranstvu. Cilj strategije treba biti animiranje svih postojećih resursa i konsolidacija trenutnih napora te ostvarenje samoodrživosti inicijative. Uz strategiju potrebno je definisati mehanizme i odgovorne za mapiranje čitavog bh. AI ekosistema, praćenje provedbe i eventualne prilagodbe strategije.

\section{Zaključak}

Kroz ovaj rad, prikazano je kako su se globalni i lokalni trendovi u oblasti umjetne inteligencije razvijali u posljednje dvije godine, osvrćući se na konkretne korake koje je Bosna i Hercegovina poduzela u pogledu obrazovanja, istraživanja i primjene AI tehnologija. Iako su neki pomaci postignuti, posebno u oblasti edukacije i digitalizacije javnih usluga, ostaje značajan prostor za dalji napredak, naročito u pogledu strateškog pristupa, etičkih regulacija i organizacije naučno-istraživačkog rada.

Bosni i Hercegovini je potrebna sveobuhvatna nacionalna AI strategija koja bi obuhvatila ne samo tehnički razvoj, već i etičke aspekte i regulatorni okvir. Institucionalna podrška, kao i razvoj interdisciplinarnih istraživačkih grupa, ključni su za uspješnu integraciju AI u sve sfere društva. Dalji koraci trebaju uključiti izgradnju specijaliziranih AI instituta, strateško upravljanje istraživačkim resursima i kontinuiranu evaluaciju rezultata. Vizija Bosne i Hercegovine u doba AI mora biti ambiciozna, ali i realna, usklađena s globalnim trendovima, kako bi zemlja bila u mogućnosti da iskoristi sve potencijale AI tehnologija u korist društva, ekonomije i budućih generacija.


\section*{Zahvala}
Autori su članovi Asocijacije za napredak nauke i tehnologije (ANNT). ANNT je nevladina organizacija koju čine mladi naučnici u oblastima prirodno-matematičkih i tehničkih nauka. Naši članovi žive i djeluju u Bosni i Hercegovini, dok dio njih radi na prestižnim svjetskim institucijama, ali su svi vezani za našu domovinu i dijele zajedničku viziju o unapređenju stanja nauke u njoj.

\section*{Biografije}

Dr. \textbf{Zlatan Ajanović} je postdoktorski istraživač na RWTH Aachen, katedri za mašinsko učenje i rezonovanje. Prije toga, bio je postdoktorski istraživač na Tehničkom Univerzitetu Delft. Bavi se istraživanjem metoda AI planiranja i učenja u robotici. Karijeru je počeo u Prevent Group u Bosni i Hercegovini gdje je radio na više projekata,  kao što je pokretanje novog proizvodnog procesa te korištenje kompjuterske vizije za osiguravanje kvalitete proizvoda. Dobitnik je najprestižnije europske stipendije za PhD studij (Marie Skłodowska-Curie Fellowship), u sklopu ITEAM projekta. Kroz ITEAM projekat je bio zaposlen kao senior istraživač u istraživačkom centru Virtual Vehicle u Grazu, te kao gostujući istraživač na Tehničkom Univerzitetu Delft, Univerzitetu u Sarajevu, AVL List i Volvo Cars. Pored toga, učestvovao je u više internacionalnih projekata sa preko 50 partnera sirom svijeta, te je aktivno učestvovao i vodio pripremanje uspješnih prijedloga projekata ukupne vrijednosti od preko 50 miliona eura. Bachelor i Master studij je završio na Elektrotehničkom fakultetu Univerziteta u Sarajevu u oblasti automatskog upravljanja. Titulu doktora tehničkih nauka stekao je na Tehničkom Univerzitetu u Grazu. Redovno objavljuje radove i prisustvuje na najprestižnijim događajima iz Robotike, Umjetne inteligencije i Automatskog upravljanja. Član je tehničkog komiteta za Inteligentna Autonomna Vozila i tehničkog komiteta za Inteligentno upravljanje pri IFAC te recenzent više časopisa i konferencija iz oblasti istraživanja. Za svoj rad, nagrađen je sa IFAC nagradom za najboljeg mladog autora te stipendijom Hans List Fond za svoju doktorsku disertaciju te DAAD AInet postdoctoral Fellowship.

M.Sc. \textbf{Hamza Merzić} je senior istraživač u oblasti umjetne inteligencije u Google DeepMind-u u Londonu (od 2018.), doktorski kandidat na University College London (od 2024.), te suosnivač Asocijacije za napredak nauke i tehnologije (ANNT). Diplomirao je na Odsjeku za automatiku i elektroniku Elektrotehničkog fakulteta Univerziteta u Sarajevu (UNSA) 2015. godine, kao najbolji student generacije i dobitnik prestižne Zlatne značke Univerziteta. Tokom bachelor studija, obnašao je dužnost potpredsjednika IEEE Studentskog ogranka u Sarajevu.
Nakon završetka osnovnih studija, upisuje magistarski studij na ETH Cirih u Švicarskoj, gdje je 2018. godine magistrirao iz oblasti robotike i umjetne inteligencije. Tokom master studija, radio je kao student asistent, a istovremeno je bio zaposlen kao istraživač u uglednoj ADRL grupi na ETH Cirih, gdje je učestvovao na projektu autonomne izgradnje u saradnji s Fakultetom za arhitekturu pri ETH. Dodatno iskustvo stekao je u industriji, radeći za Rapyuta Robotics, švicarsko-japansku kompaniju specijalizovanu za autonomnu robotiku.
Kroz rad u ANNT-u, nastoji prenijeti svoja znanja i iskustva natrag u lokalnu zajednicu, s ciljem podrške razvoju nauke i tehnologije u regiji. Aktivno objavljuje naučne radove u prestižnim časopisima i konferencijama iz oblasti umjetne inteligencije.

M.Sc. \textbf{Eldar Kurtić} je Bachelor i Master studij završio na Elektrotehničkom fakultetu Univerziteta u Sarajevu (UNSA) sa odlikovanjem Zlatna značka Univerziteta u Sarajevu na I i II ciklusu studija. Po završetku studija radio je u istraživačkom centru Robert Bosch GmbH Zentrum für Forschung und Vorausentwicklung (Renningen, Germany) na razvoju robotskih manipulatora i mašinske vizije za automatsku detekciju kvarova na industrijskim proizvodima. Nakon toga, radio je u istraživačkom centru Virtual Vehicle (Graz, Austria) na razvoju algoritama planiranja kretanja i računarske vizije za autonomna vozila. Trenutno radi kao senior istraživač i softver inženjer u oblasti mašinskog učenja i umjetne inteligencije na Institute of Science and Technology Austria (Vienna, Austria) i Neural Magic Inc. (Massachusetts, USA), sa fokusom na kompresiji dubokih neuronskih mreža. 	

Dr. med. \textbf{Bakir Kudić} je doktorant na Francis Crick Institutu i klinički istraživač u bolnici Royal Marsden u Londonu. Diplomirao je na Medicinskom fakultetu Univerziteta u Sarajevu 2022. godine kao dobitnik zlatne značke. Tokom studija bio je gostujući istraživač u Centru za proteomiku u Rijeci. Radio je kao urednik za produkciju u naučnom časopisu Biomolecules and Biomedicine, te kao asistent na Katedri za farmakologiju i toksikologiju, Medicinskog fakulteta Univerziteta u Sarajevu. Trenutno je član upravnog odbora Asocijacije za napredak nauke i tehnologije.

Dr. \textbf{Rialda Spahić} je doktorirala inženjersku kibernetiku na Norveškom Univerzitetu Nauke i Tehnologije gdje je istraživala umjetnu inteligenciju zasnovanu na riziku u primjeni autonomnog nadzora podmorskih cjevovoda. Prije doktorata, završila je računarstvo i inženjerstvo na Internacionalnom Univerzitetu u Sarajevu. Trenutno vodi implementaciju odgovorne umjetnue inteligenciju u Equinoru, globalnoj energetskoj kompaniji, gdje analizira međunarodne standarde, zakonske zahtjeve i tehnološku primjenu kako bi osigurala usklađenost AI sa etičkim principima i društvenim normama koje njeguju povjerenje i daju prioritet sigurnosti AI sistema. Pored toga, Rialda je ambasadorica i osnivačica organizacije Women in AI Norway sa volonterima u većim norveškim gradovima, organizirajući obrazovne događaje koji zatvaraju jaz između spolova i smanjuju pristrasnost u AI.

Prof. Dr. \textbf{Emina Aličković} je zaposlena kao vodeći naučnik u Eriksholm istraživačkom centru, koji je dio vodećeg svjetskog proizvođača slušnih aparata Oticon A/S, Danska. Također ima poziciju i vanrednog profesora na Odsjeku za elektrotehniku na Linkoping Univerzitetu u Švedskoj. Bachelor studije je završila na Internacionalnom Univerzitetu u Sarajevu, Odsjek za elektrotehniku i diplomirala je kao najbolji student generacije 2010 na Univerzitetu. Magistarske i doktorske studije je završila na Internacionalnom Burch Univerzitetu u Sarajevu, Odsjek za elektrotehniku i 2015. godine je stekla titulu doktora elektrotehničkih nauka. Autor je više desetina radova objavljenih u prestižnim naučnim časopisima i međunarodnim konferencijama i izumitelj je nekoliko međunarodnih patenata. Trenutno je mentor više doktoranata i postdoktoranata. Njeni istraživački interesi su statistička i adaptivna obrada signala i mašinsko učenje, sa primjenom u neuroznanosti i neurotehnologiji.

Prof. Dr. \textbf{Sead Delalić} je docent na Odsjeku za matematičke i kompjuterske nauke Prirodno-matematičkog fakulteta Univerziteta u Sarajevu, te osam godina radi kao AI inženjer i konsultant. Završio je prvi i drugi ciklus studija kao osvajač dvije zlatne značke i kao najbolji student Univerziteta u Sarajevu sa prosječnom ocjenom 10. Doktorirao je iz oblasti vještačke inteligencije i optimizacija na Univerzitetu u Sarajevu. Učestvovao je u više od 30 industrijskih i istraživačkih projekata iz oblasti vještačke inteligencije. Autor je 30 naučnih radova i koautor je jednog univerzitetskog udžbenika. Prezentovao je naučna istraživanja i stručne rezultate na konferencijama u više od 10 država. Kao član ekipe Bosne i Hercegovine, učestvovao je na nekoliko matematičkih olimpijada, kao vođa tima ili takmičar, te je više godina aktivno učestvovao u pripremama olimpijaca za međunarodna takmičenja. Trenutno je angažovan kao vođa AI/ML odjela u kompaniji Info Studio, sarađuje kao konsultant sa kompanijom Infobip, te je osnivač konsultantske kompanije Next IT.

Dr. \textbf{Suad Krilašević} je doktorirao iz oblasti teorije upravljanja na Tehničkom Univerzitetu u Delftu, Nizozemska. Rođen je u Sarajevu,  1995. godine. Završio je osnovne i master studije iz Elektronike i automatike na Elektrotehničkom fakultetu Univerziteta u Sarajevu, u septembru 2016. i 2018. godine. Njegova istraživačka interesovanja uključuju teoriju igara, multi-agentne sisteme, hibridne sisteme i traženje ekstrema. Od 2023. godine je predsjednik Asocijacije za napredak nauke i tehnologije. 

Dr. \textbf{Aida Branković} je istraživač na Australskom nacionalnom istraživačkom centru za digitalno zdravlje i spoljni saradnik na Kvinslend univerzitetu, Australija. Prije ovog imala je pozicije istraživača na Politehničkom univerzitetu u Milanu i Kvinslend univerzitetu te Zeus Spa. Dobitnik različitih nagrada od kojih su najvece Globalni Talent Award (2022) od strane australske vlade za doprinos reputaciji Australije kao svjetskog lidera u medicinskim inovacijama i Early-Career Advance Queensland Research Fellowship (AQRF) (2020). Njen rad u polju objašnjive umjetne inteligencije proglašen je najboljim naučnim radom na MedInfo konferenciji 2023. Dr Branković je vlasnik 2 patenta, 2 softverska paketa, preko 30 naučnih publikacija i 2 algoritama za kliničku podršku u odlučivanju u procesu komercijalizacije. Član je redakcije naučnog časopisa Nature Scientific Reports i član programskog odbora PRICAI. Njena istraživačka interesovanja uključuju odgovornu i pouzdanu umjetnu inteligenciju, nelinearnu identifikaciju, optimizaciju modela i njihovu primjenu u biomedicini. 

Prof. Dr. \textbf{Kenan Šehić} je uspješno odbranio doktorsku disertaciju iz oblasti računarske i primijenjene matematike na Tehničkom Univerzitetu u Danskoj 2020. godine. Nakon čega je bio na postdoktorskom istraživanju na Lund Univerzitetu pri Odsjeku za računarsku nauku. Trenutno je docent na Elektrotehničkom fakultetu Univerziteta u Sarajevu. Kao dobitnik MIT gostujuće stipendije, dio istraživačke karijere provodi na Massachusetts Institutu za Tehnologiju (MIT) pri Odsjeku za aeronautiku i astronautiku u grupi za kvantificiranje nesigurnosti. Alumni je Univerziteta u Sarajevu gdje završava prvi i drugi ciklus na Mašinskom fakultetu. Autor je 6 naučnih radova iz oblasti kvantificiranje nesigurnosti i mašinskog učenje. Njegovi interesi su kvantificiranje nesigurnosti, mašinsko učenje i optimizacija hiperparametara. Član je Asocijacije za Napredak Nauke i Tehnologije.

Prof. Dr. \textbf{Mirsad Ćosović} je vanredni profesor na Elektrotehničkom fakultetu Univerziteta u Sarajevu. Titulu doktora elektrotehnike i računarstva stekao je na Univerzitetu u Novom Sadu, radeći na projektu ADVANTAGE kao Marie Curie stipendista. Tokom studija, proveo je istraživačke posjete u Španjolskoj, Velikoj Britaniji, Slovačkoj i Kini. Autor je više radova objavljenih u prestižnim časopisima iz oblasti elektroenergetike i telekomunikacija, kao i na međunarodnim konferencijama. Trenutno radi na optimizaciji elektroenergetskih sistema i sudjeluje u projektima vezanim uz pametne mreže i zelenu energiju.

Prof. Dr. \textbf{Admir Greljo} je profesor teorijske fizike čestica i kosmologije na Univerzitetu u Bazelu, gdje od 2023. godine vodi istraživačku grupu posvećenu naprednim istraživanjima u ovim oblastima. Autor je 100 naučnih radova indeksiranih u Web of Science of kojih je 54 u kategoriji Core Collection s pripadajućim h-indeksom 31. Dodiplomski studij teorijske fizike završio je na Univerzitetu u Sarajevu, nakon čega je nastavio svoje akademsko usavršavanje na Institutu Jožef Stefan u Ljubljani, gdje je obavio doktorsko istraživanje i stekao doktorat 2014. godine na Univerzitetu u Ljubljani. Postdoktorska istraživanja je obavio na Univerzitetu u Cirihu (2014-2017), a zatim na Univerzitetu Johannesa Gutenberga u Mainzu (2017-2018). Njegov rad i značajni napreci u ovoj oblasti donijeli su mu poziciju istraživača u CERN-u (2018-2020), jednom od najvažnijih svjetskih centara za istraživanje fizike čestica. Nakon toga, Greljo je dobio prestižni SNSF Eccellenza profesorski grant (1.7 miliona CHF) na Univerzitetu u Bernu, gdje je osnovao i uspješno vodio istraživačku grupu od 2020. do 2023. godine. Danas, njegova istraživačka grupa na Univerzitetu u Bazelu istražuje fundamentalne zakone prirode, razvija teorijske modele izvan Standardnog modela, te usmjerava buduće eksperimente u fizici čestica.

Prof. Dr. \textbf{Adnan Mehonić} je vanredni profesor nanoelektronike na Univerzitetskom koledžu u Londonu (UCL), specijaliziran za memristivne i neuromorfne tehnologije. Autor je preko 100 naučnih publikacija i vlasnik 14 patenata. Kao suosnivač i tehnički direktor startupa 'Intrinsic', koji je privukao investicije u iznosu od 10 miliona dolara, njegov rad je usmjeren na razvoj novih memorijshih tehnologija i energetski efikasnih računarskih arhitektura. Dr. Mehonić je također direktor master programa iz nanotehnologije i glavni urednik časopisa APL Machine Learning.

\section*{References}

\medskip

Ahn, Michael, Anthony Brohan, Noah Brown, Yevgen Chebotar, Omar Cortes, Byron David, Chelsea Finn, et al. 2022. “Do As I Can, Not As I Say: Grounding Language in Robotic Affordances.” In . arXiv.

Ajanović, Zlatan, Emina Aličković, Aida Branković, Sead Delalić, Eldar Kurtić, Salem Malikić, Adnan Mehonić, Hamza Merzić, Kenan Šehić, and Bahrudin Trbalić. 2022. “Vision for Bosnia and Herzegovina in Artificial Intelligence Age: Global Trends, Potential Opportunities, Selected Use-Cases and Realistic Goals.” In Scientific-Professional Conference “Artificial Intelligence in Bosnia and Herzegovina”- Research, Application and Development Perspectives, 13–46. Federalno ministarstvo obrazovanja i nauke/znanosti: Fondacija za inovacijski i tehnološki razvoj.

Alonso, Cristian, Andrew Berg, Siddharth Kothari, Chris Papageorgiou, and Sidra Rehman. 2022. “Will the AI Revolution Cause a Great Divergence?” Journal of Monetary Economics 127 (April):18–37.

Anthropic. 2024. “Introducing the next Generation of Claude.” 2024..
Chiang, Wei-Lin, Lianmin Zheng, Ying Sheng, Anastasios Nikolas Angelopoulos, Tianle Li, Dacheng Li, Hao Zhang, et al. 2024. “Chatbot Arena: An Open Platform for Evaluating LLMs by Human Preference.” arXiv.

“Core.Edu.Au - ICORE Rankings Portal.” n.d. Accessed September 8, 2024..
Davis, Wes. 2024. “All the News about SB 1047, California’s Bid to Govern AI.” The Verge. September 3, 2024.

DeepSeek-AI, Aixin Liu, Bei Feng, Bin Wang, Bingxuan Wang, Bo Liu, Chenggang Zhao, et al. 2024. “DeepSeek-V2: A Strong, Economical, and Efficient Mixture-of-Experts Language Model.” arXiv..
Dubey, Abhimanyu, Abhinav Jauhri, Abhinav Pandey, Abhishek Kadian, Ahmad Al-Dahle, Aiesha Letman, Akhil Mathur, et al. 2024. “The Llama 3 Herd of Models.” arXiv.

EU. 2024. Regulation (EU) 2024/1689 of the European Parliament and of the Council of 13 June 2024 Laying down Harmonised Rules on Artificial Intelligence and Amending Regulations (EC) No 300/2008, (EU) No 167/2013, (EU) No 168/2013, (EU) 2018/858, (EU) 2018/1139 and (EU) 2019/2144 and Directives 2014/90/EU, (EU) 2016/797 and (EU) 2020/1828 (Artificial Intelligence Act) (Text with EEA Relevance).

FFG. 2024. “Das große COMET-Netzwerk | FFG.” 2024.

Gemini Team, Rohan Anil, Sebastian Borgeaud, Jean-Baptiste Alayrac, Jiahui Yu, Radu Soricut, Johan Schalkwyk, et al. 2024. “Gemini: A Family of Highly Capable Multimodal Models.” arXiv..
Gemini Team, Petko Georgiev, Ving Ian Lei, Ryan Burnell, Libin Bai, Anmol Gulati, Garrett Tanzer, et al. 2024. “Gemini 1.5: Unlocking Multimodal Understanding across Millions of Tokens of Context.” arXiv.

Gertner, Jon. 2012. The Idea Factory: Bell Labs and the Great Age of American Innovation. The Idea Factory: Bell Labs and the Great Age of American Innovation. New York, NY, US: Penguin Press.

Gholami, Amir, Sehoon Kim, Zhen Dong, Zhewei Yao, Michael W. Mahoney, and Kurt Keutzer. 2022. “A Survey of Quantization Methods for Efficient Neural Network Inference.” In Low-Power Computer Vision. Chapman and Hall/CRC.

Google Scholar. 2024. “Artificial Intelligence - Google Scholar Metrics.” 2024..
Harari, Haim, Olaf Kuebler, and Markl. 2006. “Recommended Steps towards the Establishment of the Institute of Science and Technology – Austria (ISTA).”
Hinton, Geoffrey, Oriol Vinyals, and Jeff Dean. 2015. “Distilling the Knowledge in a Neural Network.” arXiv.

HRZZ. 2009. “Dokumenti – HRZZ.” 2009.

IDDEEA. 2024a. “IDDEEA BiH učestvuje u pilot projektu Evropske unije za jačanje integriteta i tačnosti izbornog procesa BiH – IDDEEA.” 2024.

———. 2024b. “Kvalifikovani elektronski potpis dostupan građanima od jula 2024.godine – IDDEEA.” 2024.

Internacionalni Univerzitet u Sarajevu. 2023. “Curricula.” 2023..
International Telecommunication Union. 2023. “Bosnia and Herzegovina Digital Development Country Profile.”

ISTA. 2019. “IST Austria No. 3 among the World’s Best Research Institutions.” Institute of Science and Technology Austria (ISTA) (blog). 2019.

IVI. 2024. “Projekti.” Istraživačko-razvojni institut za veštačku inteligenciju (blog). 2024..
Jiang, Albert Q., Alexandre Sablayrolles, Arthur Mensch, Chris Bamford, Devendra Singh Chaplot, Diego de las Casas, Florian Bressand, et al. 2023. “Mistral 7B.” arXiv.

Jumper, John, Richard Evans, Alexander Pritzel, Tim Green, Michael Figurnov, Olaf Ronneberger, Kathryn Tunyasuvunakool, et al. 2021. “Highly Accurate Protein Structure Prediction with AlphaFold.” Nature 596 (7873): 583–89.

Laurençon, Hugo, Léo Tronchon, Matthieu Cord, and Victor Sanh. 2024. “What Matters When Building Vision-Language Models?” arXiv..
Luijkx, Jelle, Zlatan Ajanovic, Laura Ferranti, and Jens Kober. 2022. “PARTNR: Pick and Place Ambiguity Resolving by Trustworthy iNteractive leaRning.” In NeurIPS 2022 Workshop on Robot Learning. arXiv.

Ma, Runyu, Jelle Luijkx, Zlatan Ajanovic, and Jens Kober. 2024. “ExploRLLM: Guiding Exploration in Reinforcement Learning with Large Language Models.” arXiv.

Mehonic, A., and A. J. Kenyon. 2022. “Brain-Inspired Computing Needs a Master Plan.” Nature 604 (7905): 255–60.

Mehonic, Adnan, and Jason Eshraghian. 2023. “Brains and Bytes: Trends in Neuromorphic Technology.” APL Machine Learning 1 (2): 020401.

Mehonic, Adnan, Abu Sebastian, Bipin Rajendran, Osvaldo Simeone, Eleni Vasilaki, and Anthony J. Kenyon. 2020. “Memristors—From In-Memory Computing, Deep Learning Acceleration, and Spiking Neural Networks to the Future of Neuromorphic and Bio-Inspired Computing.” Advanced Intelligent Systems 2 (11): 2000085.

Mnih, Volodymyr, Koray Kavukcuoglu, David Silver, Andrei A. Rusu, Joel Veness, Marc G. Bellemare, Alex Graves, et al. 2015. “Human-Level Control through Deep Reinforcement Learning.” Nature 518 (7540): 529–33.

Mügge, Daniel. 2024. “EU AI Sovereignty: For Whom, to What End, and to Whose Benefit?” Journal of European Public Policy 31 (8): 2200–2225.

NIST, NIST. 2023. “Artificial Intelligence Risk Management Framework (AI RMF 1.0).” National Institute of Standards \& Technology, USA.

OpenAI, Josh Achiam, Steven Adler, Sandhini Agarwal, Lama Ahmad, Ilge Akkaya, Florencia Leoni Aleman, et al. 2024. “GPT-4 Technical Report.” arXiv.

Our World in Data. 2024a. “Number of R\&D Researchers per Million People.” Our World in Data. 2024.

———. 2024b. “Research \& Development Spending as a Share of GDP.” Our World in Data. 2024..
Ouyang, Long, Jeffrey Wu, Xu Jiang, Diogo Almeida, Carroll Wainwright, Pamela Mishkin, Chong Zhang, et al. 2022. “Training Language Models to Follow Instructions with Human Feedback.” Advances in Neural Information Processing Systems 35 (December):27730–44.

Rafailov, Rafael, Archit Sharma, Eric Mitchell, Christopher D. Manning, Stefano Ermon, and Chelsea Finn. 2023. “Direct Preference Optimization: Your Language Model Is Secretly a Reward Model.” Advances in Neural Information Processing Systems 36 (December):53728–41.

Republika Srbija. 2018. “Zakon o fondu za nauku Republike Srbije.” 2018.

Schulman, John, Filip Wolski, Prafulla Dhariwal, Alec Radford, and Oleg Klimov. 2017. “Proximal Policy Optimization Algorithms.” arXiv.

Silver, David, Aja Huang, Chris J. Maddison, Arthur Guez, Laurent Sifre, George van den Driessche, Julian Schrittwieser, et al. 2016. “Mastering the Game of Go with Deep Neural Networks and Tree Search.” Nature 529 (7587): 484–89.

SJR. 2024. “SJR - International Science Ranking.” 2024.

Univerzitet u Sarajevu. 2024. “DSAI \& ETF - Data Science and Artificial Intelligence.” 2024.

Vaswani, Ashish, Noam Shazeer, Niki Parmar, Jakob Uszkoreit, Llion Jones, Aidan N. Gomez, Lukasz Kaiser, and Illia Polosukhin. 2023. “Attention Is All You Need.” arXiv.

Vecernji. 2024. “BiH po brzini interneta zaostaje za većinom zemalja u Europi, a u regiji ima brži internet samo od RH.” 2024.

Vlada Republike Srbije. 2021. “Odluka o Osnivanju Istraživačko-Razvojnog Instituta za Veštačku Inteligenciju Srbije: 24/2021-35, 38/2021-24 (Ispravka).” 2021.

Wang, Yubo, Xueguang Ma, Ge Zhang, Yuansheng Ni, Abhranil Chandra, Shiguang Guo, Weiming Ren, et al. 2024. “MMLU-Pro: A More Robust and Challenging Multi-Task Language Understanding Benchmark.” arXiv.

Yang, An, Baosong Yang, Binyuan Hui, Bo Zheng, Bowen Yu, Chang Zhou, Chengpeng Li, et al. 2024. “Qwen2 Technical Report.” arXiv.

Kisačanin, Branislav, Nebojša Vasiljević, and Dubravko Ćulibrk. 2021. “Elaborat o Opravdanosti Osnivanja Istraživačko-Razvojnog Instituta za Veštačku Inteligenciju Srbije.”

\end{document}